\documentstyle[11pt,epsfig]{article}

\def\numunux{\nu_\mu\rightarrow \nu_x}
\def\Journal#1#2#3#4{{#1} {\bf #2}, #3 (#4)}

\newlength{\dinwidth}
\newlength{\dinmargin}
\def\etal{{\it et\ al.}}

\def\NIMA{{\em Nucl. Instrum. Methods} A}

\def\PLB{{\em Phys. Lett.}  B}

\def\PRD{{\em Phys. Rev.} D}

\begin{document}

\begin{center}
{\large \bf A Test of $\numunux$ Oscillations and determination
of the $\Delta m^2$ parameter of \\ Atmospheric
Neutrinos using the $L/E$ method}
\end{center}

\begin{center}
G. Mannocchi$~^1$, L. Periale$~^1$, P. Picchi$~^{1,2}$,
F. Pietropaolo$~^3$, S. Ragazzi$~^4$, A. Rubbia$~^{5,6}$\\
~\\
{\em
$~^1$ Instituto di 
Cosmogeofisica del C.N.R., 
corso Fiume 4, Torino, Italy \\
$~^2$ Lab. Naz. di Frascati dell'INFN, 
via E. Fermi 40, Frascati (Roma), Italy \\
 
$~^3$ INFN, Sezione di Padova, via Marzolo 8, Padova, Italy \\

$~^4$ Dipartimento di Fisica dell'Universita' and INFN, 
via Celoria 16, Milano, Italy \\

$~^5$ Swiss Federal Institute of Technology Zurich, 
CH-8092 Zurich, Switzerland\\
$~^6$ CERN, PPE Division, 
CH-1211 Geneva 23, Switzerland \\
}

\end{center}

\centerline{\today}

\abstract{We point out that the measurement of the $L/E$ distribution
of atmospheric neutrino events with sufficient resolution
in order to observe the characteristic modulations
of neutrino oscillations would provide a unique way to test
the $\numunux$ oscillations nature of the atmospheric
neutrino anomaly\cite{kamatmo} in the parameter range
$5\times 10^{-4} < \Delta m^2 < 2\times 10^{-3}\rm\ eV^2$. 
With an integrated exposure of a few tens of kiloton-year,
this method can cover the lower part of the $\Delta m^2$ 
region of the currently allowed atmospheric 
neutrino solution\cite{superkamiokande} and is therefore
complementary to the one explorable with planned long-baseline
neutrino beam experiments (e.g. Refs. \cite{icarus,icarusprop}). 
We stress that experimentally
this method relies on the ability to measure high energy muons
($E_\mu>5\rm\ GeV$). A large magnetized calorimeter of several kilotons
would be ideally suited for this measurement.
The existing LEP detectors augmented with atmospheric
neutrino triggers could also be used to perform such a study.}

\section{Introduction}
Recent SuperKamiokande data show an asymmetry in the rate of upward and 
downward atmospheric muon-like events\cite{kamatmo,superkamiokande}. 
In order to verify that this 
phenomenon is really due to neutrino oscillations, an effective method 
consists in observing the $\sin^2(1.27 \Delta m^2 L/E)$ modulation characteristic of 
a given $\Delta m^2$ when the event rate is plotted as a function of the
reconstructed $L/E$ of the events. 
Such a method, capable of measuring $\Delta m^2$ in the range between 
$5\times 10^{-4}$ and $2\times 10^{-3}\rm\ eV^2$
exploiting atmospheric neutrino events, has been advocated in a 
recent note\cite{PioFrancesco}. 
The method has the advantage of being
practically insensitive to the precise knowledge of 
the atmospheric neutrino flux since the oscillation pattern is found by 
striking dips in the $L/E$ distribution
while the neutrino interaction spectrum is known to be a 
slowly varying function of $L/E$.
It was shown that the main experimental requirement is that the error on $L/E$ 
is the following:
\begin{equation}\label{formu}
\delta(L/E) \ll \pi/(1.27 \Delta m^2)
\end{equation}
This implies that precise measurements of energy and direction of both the muon 
and hadrons are needed in order to reconstruct precisely the neutrino $L/E$. 
Note that all past and present large scale underground atmospheric 
neutrino experiments are
unable to determine accurately the momentum of high 
energy muons (e.g. $E_\mu > 5\ \rm GeV$).  
This implies that the capability to study the $L/E$ distribution is currently
quite limited.
The ability to measure high momentum muons 
requires the construction of a magnetized
calorimeter with a mass of several kilotons in order
to have enough of an event interaction rate.
\par
In section \ref{optimz}, we define a set of selection cuts that
improve the reconstruction of the $L/E$ distribution of events
which do not strongly rely on the hadronic energy resolution.
In section \ref{osc}, we show how the oscillation can be
found and the oscillation parameters determined in a generic
magnetized calorimeter.
In section \ref{lepnus}, we argue
that the existing LEP detectors --- able to measure muon momentum 
and direction with a coarse hadronic energy resolution --- could 
fulfill the required criteria to observe the $L/E$ modulations.

\begin{table}[b]
\begin{center}
\begin{tabular}{lccc}
\hline
Selection & $\nu_\mu$ & $\bar\nu_\mu$ & $\delta(L/E)$ \\
          & Fraction(\%) & Fraction(\%) & RMS (Km/GeV) \\
\hline
All $(E_\nu>1\rm\ GeV)$ & 100 & 100 & 765 \\
$E_\mu>2.5\rm\ GeV$ & 31.7 & 45.1 & 560 \\
$E_{had} < 0.5 E_\mu$ & 13.7 & 27.0 & 350 \\
$\theta<75^o$ or $\theta>105^o$ & 11.2 & 21.9 & 345 \\
\hline
\end{tabular}
\caption{Neutrino and anti-neutrino event selections.}
\end{center}
\label{tab:sele}
\end{table}

\section{Optimizing the $L/E$ resolution}\label{optimz}
The atmospheric neutrino 
interaction energy spectrum for $E > 1\ \rm GeV$ has roughly a
$E^{-1.7}$ dependence\cite{Lipari}.
The basic idea to satisfy Eq.~\ref{formu}
consists in selecting only the neutrino events where 
the muon direction is close to that of the incoming neutrino. 
Given the energy dependence of the atmospheric neutrinos,
the events with high energy muons are more likely to fulfill the condition.
Therefore, there is a cut on the muon momentum for which
the total event rate and the $L/E$ resolution are optimized.
Moreover the effective cut in the neutrino energy 
removes the smearing due to the Fermi motion of the target nucleon. 
\par
On a sample of simulated atmospheric neutrino interactions,
we placed a cut on the muon energy at 2.5 GeV to select a sample of 
events where an 80\% fraction has the opening angle 
smaller that $10^o$. The remaining 20\% is mainly due to high-$y$ high 
neutrino energy events
and is small due to the $E^{-1.7}$ dependence of the neutrino 
interaction spectrum. It can be further reduced requiring small hadronic energy 
deposition; a cut of $E_{had} < 0.5\times E_\mu$ satisfies the above requirements. 
We point out 
that for this purpose a good hadronic energy resolution is not needed.
The neutrino energy is reconstructed using both muon and hadron energy but
the neutrino energy resolution is only partially affected by the rough 
hadronic energy resolution because of the applied selection criteria. 
\par
Another important cut is applied on the incoming neutrino zenith angle $\theta$; 
in order to minimize the error on $L$, only the events with 
$\theta<75^o$ or $\theta>105^o$ should be retained (see Ref.~\cite{PioFrancesco})
to well define the sets of upward and downward events.
The selection efficiency of this method is close to 15 \% for events in the 
detector with neutrino energy above 1.0 GeV. The selection efficiencies
for neutrinos and antineutrinos are summarized in Table~\ref{tab:sele}.
The efficiency for antineutrinos is higher due to
$d\sigma/dy\propto(1-y)^2$ compared to the flat $d\sigma/dy$ for
neutrinos.

\section{Oscillation parameters determination}\label{osc}
To illustrate the method, we consider a generic calorimetric detector with
muon identification and measurement capability. The assumed resolutions
are listed in Table~\ref{tab:spinetone}.

\begin{table}[tb]
\begin{center}
\begin{tabular}{lcc}
\hline
                &       Muons   &    Hadrons       \\
\hline
&\\
{\em Generic calorimeter:}\\
$\delta E/E$    &      5\%      &  $80\%/\sqrt{E}$  \\
$\delta\theta$  &   10 mrad     &      not used    \\
&\\
\hline
\end{tabular}
\caption{Resolutions assumed for a generic calorimetric detector
with muon identification and measurement capability.}
\label{tab:spinetone}
\end{center}
\end{table}

Given the oscillation parameter $\Delta m^2$ of interest,
we expect that the downward atmospheric neutrinos above 2.5 GeV will not 
have oscillated while the upward going neutrinos will show the 
oscillation\cite{superkamiokande}. 
The sensitivity to oscillations is therefore obtained by comparing the measured 
L/E distributions of upward and downward going neutrino events. 
For upward going events, the distance L is obtained from the $\cos\theta$
with the
expression $L=\sqrt{R^2-(R-d)^2\sin^2\theta}-(R-d)\cos\theta$
where $R$ is approximately the earth radius and $d$ is the depth of the atmosphere. 
For downward going events, the mirror-distance 
L' is calculated replacing in the expression $\theta \rightarrow -\theta$, 
in order obtain an 
``unoscillated'' upward going $L'/E$ spectrum. This method only relies on 
the fact that the angular distribution of atmospheric neutrinos is 
symmetric in $\cos\theta$ for energies above 5 GeV and below 5 GeV
the geomagnetic effect and the solar activity perturbances are
small\cite{Lipari}. In the considered $\cos\theta$ region (see section
\ref{optimz}), the geomagnetic correction is even smaller.
The $\numunux$ 
conversions are extracted by searching for the characteristic 
pattern of oscillations in the $L/E$ distribution normalized by 
the $L'/E$ distribution ($R = (L/E)/(L'/E)$).

\begin{figure}[tb]
\begin{center}
\mbox{\epsfig{file=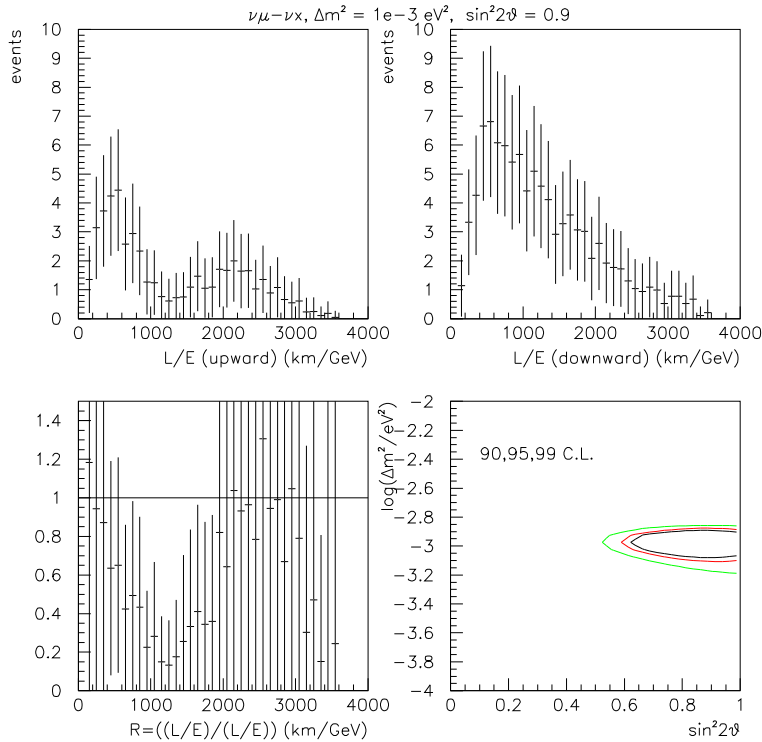,height=11cm}}
\vspace{-1.0cm}
\end{center}
\caption{Measured $L/E$ distribution in presence
of $\numunux$ oscillations with parameters
$\Delta m^2 = 1\times 10^{-3}$ and $\sin^2 2\theta=0.9$
after an exposure of 30 kt-year
for a) upward going muons; b) downward
going muons; c) the ratio $R$; d) the confidence levels for the
determination of the mass and mixing angle.}
\label{fig:one}
\end{figure}

\begin{figure}[tb]
\begin{center}
\mbox{\epsfig{file=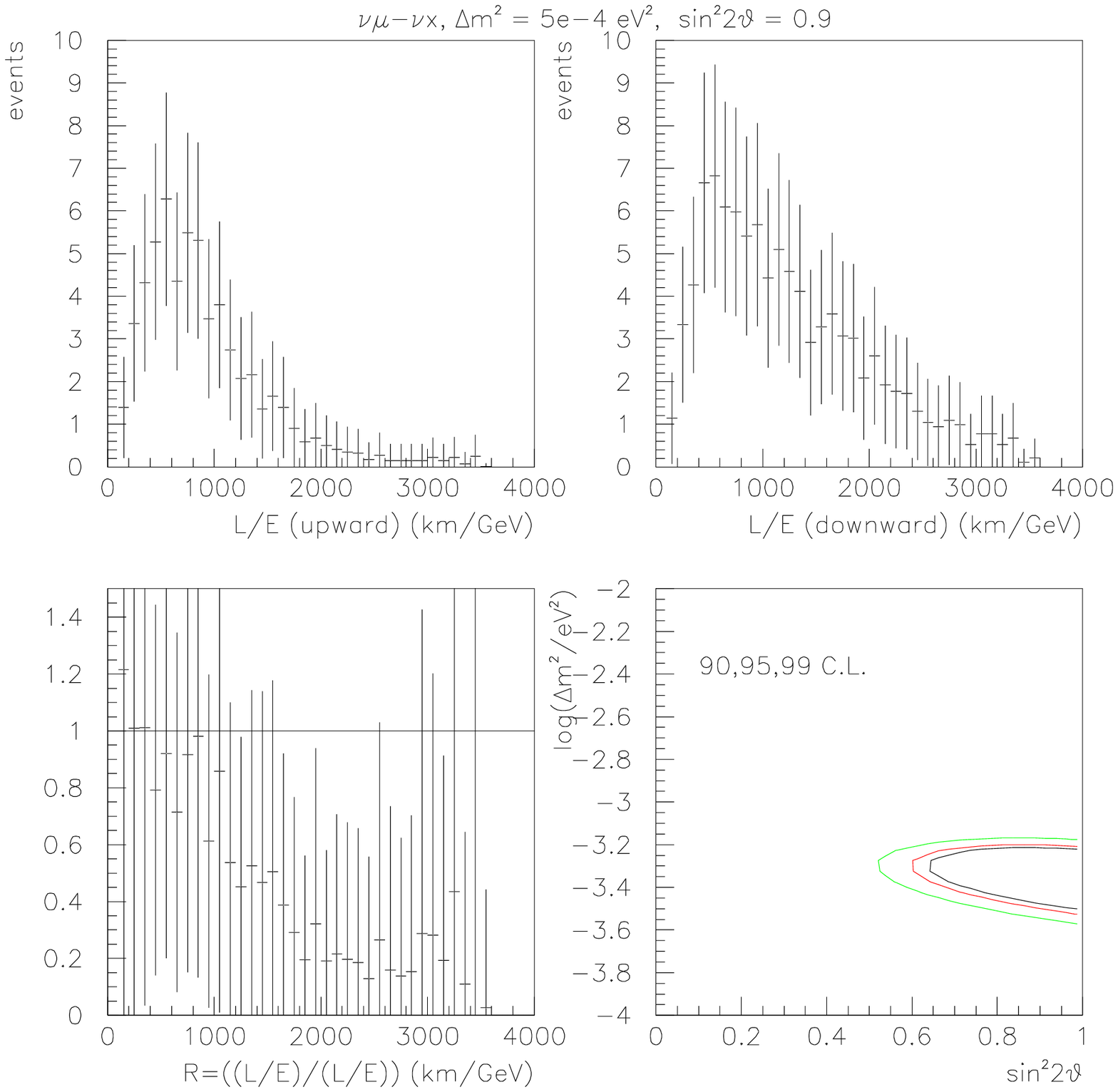,height=11cm}}
\vspace{-1.0cm}
\end{center}
\caption{Measured $L/E$ distribution in presence
of $\numunux$ oscillations with parameters
$\Delta m^2 = 5\times 10^{-4}$ and $\sin^2 2\theta=0.9$
after an exposure of 30 kt-year
for a) upward going muons; b) downward
going muons; c) the ratio $R$; d) the confidence levels for the
determination of the mass and mixing angle.}
\label{fig:two}
\end{figure}

In Figures \ref{fig:one},\ref{fig:two} 
we show the measured $L/E$ distribution in presence
of $\numunux$ oscillations with parameters
$\Delta m^2 = 1\times 10^{-3}$ and $\Delta m^2 = 5\times 10^{-4}$ 
and $\sin^2 2\theta=0.9$
after an exposure of 30 kt-year
for upward going and downward going muons. 
The oscillation parameters are extracted by comparing
the two distributions. 
For simplicity, we show the ratio $R$ of the two distributions 
where we have assumed symmetric errors.
The position of the first minimum in the R distribution 
(corresponding to an $(L/E)_{min}$) yields an estimate of the oscillation 
parameter $\Delta m^2$ through the oscillation formula. 
The best estimate of 
the parameters in the ($\Delta m^2$, $\sin^22\theta$) plane are obtained by fitting 
the oscillation formula to the observed R distribution.
The confidence levels (see Figures \ref{fig:one},\ref{fig:two}) 
for the determination of the mass difference
and mixing angle at the 90, 95 and 99\% shows the
power of this method.

\section{Atmospheric Neutrinos in LEP detectors}\label{lepnus}
The existing LEP detectors (ALEPH\cite{ALEPH}, DELPHI\cite{DELPHI}, L3\cite{L3}
and OPAL\cite{OPAL}), 
equipped with some additional apparata, could fulfill the required criteria; namely:
\begin{itemize}
\item large mass (several kt);
\item magnetized muon spectrometer;
\item segmented hadron calorimeter;
\item very precise tracking chambers (TPC's and L3 MUCH).
\end{itemize}
We note that for our purpose the 
three detectors ALEPH, DELPHI and OPAL have very similar characteristics
with most of their mass concentrated in their hadronic calorimeters acting 
as return yokes for their inner magnetic field. The L3 detector has a large 
mass in its iron magnet surrounding completely the large precise muon chambers 
and the inner sub-detectors. 
We estimate that a useful mass of $\approx 3\ \rm kt$ 
can be obtained with the iron of the 
L3 detector magnet; the hadron calorimeters of ALEPH, OPAL and DELPHI amount to 9~kt;
about one more kt can be added if one includes the electromagnetic calorimeters.
\par
For the events occurring in the hadronic calorimeters 
the neutrino energy is estimated as 
the total visible energy and an appropriate cut on the hadronic activity 
will be used to improve the correlation between neutrino and muon direction.
We stress the importance of the magnetic field available in all LEP detectors in 
order to obtain a reliable muon momentum measurement:
\begin{itemize}
\item In the case of ALEPH, the acceptance for the muon to traverse the 
inner tracking detector is large since the hadron calorimeter
was built to completely cover the inner tracker.
In this case a very precise determination of the muon momentum is 
available; for the events with this topology the measurement error 
is dominated by the error of the backward extrapolation into the hadron
calorimeter. For the muons that do not cross the inner tracker,
the fine segmentation of the hadron
calorimeter should permit the momentum to be measured 
by curvature (using the residual magnetic field) or by range-out. 
\item For the events in the L3 magnet, the hadronic energy cannot be 
reliably estimated. Nevertheless, we recall that the cut on muon momentum 
suppresses topologies with large hadronic energy and a further 
reduction can be obtained by requiring no other activity that the muon 
in the muon chambers. So we expect that the muon momentum vector will 
be a good estimator of the incoming neutrino energy and direction.
To efficiently veto cosmic ray muon backgrounds, we expect that the L3 magnet
will have to be covered with active large area detectors (e.g. RPC's).
\end{itemize}
\par
We deduce from the technical reports of the LEP experiments\cite{ALEPH,
DELPHI,L3,OPAL} that the resolutions for reconstructing the events
in the topologies described above are well matched with the one
hypothesized for the generic calorimeter of section~\ref{osc}. A more
precise description and discussion on the use of the LEP detectors will
be given after detailed simulations will have been performed.


\section{Conclusions}
In this note we have shown that the neutrino oscillation 
parameter, $\Delta m^2$, can be precisely determined through the 
measure of atmospheric neutrino flux as a function of 
$L/E$ in the search region between $5\times 10^{-4}$ and 
$2\times 10^{-3}\rm\ eV^2$ even 
with a large mass detector with rough hadronic resolution in the GeV range.
The four LEP detectors have a total mass of about 10 kt and
resolutions matching the requirements to use the $L/E$ method.
It could be very interesting to explore their capabilities as
atmospheric neutrino detectors.

\section*{Acknowledgments}
We thank Paolo Lipari for providing us with the atmospheric
neutrino spectra.


\begin{thebibliography}{000}

\bibitem{kamatmo} K.S. Hirata et al., 
\Journal{\PLB}{205}{416}{1988}; 
\Journal{\PLB}{280}{146}{1992};
Y. Fukuda et al., \Journal{\PLB}{335}{237}{1994}.

\bibitem{superkamiokande} Y.~Totsuka, (SuperKamiokande Collab.),
in LP'97, International Lepton-Photon Symposium,
Hamburg, Germany, 1997 to appear in the Proceedings. 
M.~Nakahata, (SuperKamiokande Collab.),
in HEP'97, International Europhysics Conference in High Energy
Physics, Jerusalem, Israel, 1997, to appear in the Proceedings. 

\bibitem{icarus} ICARUS Collab., Laboratori Nazionali del Gran Sasso
(LNGS) Int. Note, LNGS - 94/99 (Vols I-II), unpublished; LNGS 95/10, unpublished.
P. Benetti \etal, \Journal{\NIMA}{327}{173}{1993};\Journal{\NIMA}{332}{395}{1993};
P. Cennini \etal, \Journal{\NIMA}{333}{567}{1993};\Journal{\NIMA}{345}{230}{1994};
\Journal{\NIMA}{355}{660}{1995}.

\bibitem{icarusprop} 
ICARUS-CERN-Milano~Collab., CERN/SPSLC 96-58, SPSLC/P 304, December 1996;
J. P. Revol \etal, ICARUS-TM-97/01, 5 March 1997, unpublished.

\bibitem{PioFrancesco} P. Picchi and F. Pietropaolo, ``Atmospheric Neutrino Oscillations Experiments'', ICGF Rap. Int. 344/1997,
 Torino, 1997 (CERN preprint SCAN-9710037).

\bibitem{Lipari} V. Agrawal \etal, \Journal{\PRD}{53}{1314}{1996}; 
P. Lipari, private communication.

\bibitem{ALEPH} ALEPH Collab., ``The ALEPH handbook 1995'', Vol I,
ISBN 92-9083-072-7 (CERN).

\bibitem{DELPHI} DELPHI Collab., ``DELPHI progress report'',
CERN/LEPC 84-16, LEPC/PR 6, September 1984.

\bibitem{L3} L3 Collab., ``The construction of the L3 experiment'',
L3 Preprint \#000, October 1989 (unpublished).

\bibitem{OPAL} OPAL Collab., \Journal{\NIMA}{305}{275}{1991}.

\end{thebibliography}
\end{document}